# Magnetic phase transition and magnetocaloric effect in $PrCo_9Si_4$ and $NdCo_9Si_4$

**Niharika Mohapatra and E.V. Sampathkumaran**
*Tata Institute of Fundamental Research, Homi Bhabha Road, Mumbai 400005, India.*

The compounds, $PrCo_9Si_4$ and $NdCo_9Si_4$, have been recently reported to exhibit first-order ferromagnetic transitions near 24 K. We have subjected this compound for further characterization by magnetization, heat-capacity and electrical resistivity ($\rho$) measurements at low temperatures in the presence of magnetic fields, particularly to probe magnetocaloric effect and magnetoresistance. The compounds are found to exhibit rather modest magnetocaloric effect at low temperatures peaking at Curie temperature, tracking the behavior of magnetoresistance. The magnetic transition does not appear to be first order in its character.

1. **Introduction**

The compounds derived from LaCo$_{13}$ attracted some attention due to their potential for permanent magnet materials [1]. This compound crystallizes in NaZn$_{13}$-type cubic structure. However, corresponding Pr and Nd analogues do not seem to form, but a partial replacement of Co by Si could stabilize this structure in a narrow concentration regime, typically x<3.0 for RCo$_{13-x}$Si$_x$ (R= rare-earths). Further replacement of Co by Si results in a tetragonal distortion (Space group: *I4/mcm*) [2-5]. Here we focus on the compounds with the 1:9:4 composition. The Y compound with this stoichiometry [6] exhibits itinerant electron ferromagnetism below 25 K, while La compound shows itinerant electron metamagnetism [7]. The Ce analogue has been reported to be an intermediate-valent compound [8]. The most recent report [9] proposes that the compounds, PrCo$_9$Si$_4$ and NdCo$_9$Si$_4$, exhibit first-order ferromagnetic transitions near ($T_C$=) 24 and 12 K respectively. Clearly, this family of compounds thus exhibits a variety of interesting magnetic properties. In this article, we focus on these Pr and Nd compounds, as the materials with first-order magnetic transitions are believed to be potential candidates to search for large magnetocaloric effect (MCE) [10]. It may be noted that many Fe-based compounds of this family have been studied from MCE angle [10], whereas such searches on Co-based systems are scarce in the literature. With this primary motivation, we have carried out electrical resistivity ($\rho$), isothermal magnetization (M), and heat-capacity (C) measurements at low temperatures in the presence of magnetic fields, the results of which are presented in this article.

2. **Experimental details**

The samples in the polycrystalline form were prepared by arc melting stoichiometric amounts of constituent elements in an atmosphere of argon [Pr (>99.9%), Nd (>99.9%), Co (>99.9%) and Si (>99.999%) were procured from Leico industries]. The ingots were melted four times. The weight loss after arc melting was not more than 1%. Subsequently, the ingots were wrapped in Ta foils and annealed at 1050 C for about a week in an evacuated sealed quartz tube and furnace-cooled. The specimens are not brittle and are found to be stable in air. The samples are found to be essentially single phase (tetragonal form) by x-ray diffraction. The lattice constants obtained by least squares fitting of the d-spacings are: Pr, *a*= 7.789 ± 0.004 Å, *c*= 11.513 ± 0.006 Å; Nd, *a*= 7.788 ± 0.004 Å, *c*= 11.509 ± 0.006 Å. The specimens were further characterized by scanning electron microscope and found to be homogeneous; the composition obtained by energy dispersive x-ray analysis taken on grains reveal that the composition essentially corresponds to ideal stoichiometry 1:9:4. The $\rho$ measurements (2-300 K) and C measurements (2-50 K) as a function of temperature (T) in the presence of few selected fields were performed at low temperatures employing a commercial physical property measurements system (Quantum Design, USA). (Since there are microcracks in the samples, the absolute values of $\rho$ reported here are overestimated). Dc magnetization (M) studies (2 – 300 K) were carried out employing a commercial vibrating sample magnetometer (Oxford Instruments, UK).

## 3. Results and discussion

In order to understand magnetic ordering behavior of these compounds, the results of ρ and C measurements are shown in the low temperature range only in figure 1 and 2. *We first look at the data for the Pr compound*: As in Ref. 9, in the ac χ(T) plot (not shown here to minimize repetitions of the literature data), as the temperature (T) is lowered, there is a double-peak structure, one near 24 K and another peak near 18 K. We attribute the 23-24K feature to the magnetism from Co, considering that Y analogue also shows magnetic transitions nearly at the same temperatures [6]. The feature near 18 K might arise from complex spin reorientation effects. We did not observe any frequency dependence of the peak temperature in the entire temperature interval, thereby ruling out spin-glass freezing. In figure 1a, we see corresponding features due to magnetic transitions in the C(T) data as well in zero field: There is an upturn near 24 K and a broad shoulder below 18 K. These features are more clearly visible in the plot of C/T (Fig. 1b). We have also obtained the ρ data in cooling as well as warming cycles. There is a sudden fall of ρ near 24 K due to the loss of spin-disorder contribution. The variation of ρ across 18 K is rather smooth, thereby implying that the loss of spin-disorder contribution is so gradual that the 18K-transition is not clearly resolved in the ρ data. These features are in good agreement in with Ref. 8. *With respect to NdCo$_9$Si$_4$,* in zero magnetic field, there is a drop in ρ near 22 K with a corresponding anomaly in C (Fig. 2), indicating the onset of magnetic ordering. There are additional shoulders in C and C/T at a lower temperature (around 10 K). The value of $T_C$, while higher than that reported in Ref. 9, is in fair agreement with Ref. 3. We have also studied different specimens for this compound and we noticed that the value of $T_C$ is sample-dependent varying in the range 20 to 30 K. It is not clear whether this is due to different degree of disorder. As far as this article is concerned, we stick to the specimen with a $T_C$ of 22 K. As mentioned for the Pr compound, this transition may arise from Co ions.

*We now discuss the magnetocaloric effect behavior.*
**PrCo$_9$Si$_4$:**

In figure 3a, we show typical isothermal field-dependence of M measured at close intervals (about 3 K) of temperature (with the primary aim of inferring MCE behavior). For the sake of clarity, we show the data at selected temperatures only. M sharply rises for initial applications of magnetic field below 25 K with a tendency for saturation at high fields. Hysteretic behavior of M is not observed thereby indicating that this compound is a soft ferromagnet. The plot of $M^2$ versus H/M (Arrott plot, Fig. 3b) reveals a spontaneous moment below 25 K. The extrapolated saturation moment , say at 1.8 K (about 3.3 $\mu_B$/per formula unit) is nearly twice of that observed [6] for YCo$_9$Si$_4$, which suggests that Pr is also possibly magnetically ordered, though it is difficult to pinpoint $T_C$ for Pr ion. MCE is measured by the isothermal entropy change [ΔS= S(H)-S(0)] upon application of a magnetic field. This can be determined using the well-known thermodynamic Maxwell relation [10] connecting S, M, T and H. The values of ΔS thus obtained from the magnetization data are shown in figure 4a typically for few field variations. As suggested in Ref. 10, we have presented the data into the units of mJ/ccK, rather than J/molK (theoretical density: 7.45). The sign of ΔS is negative consistent with ferromagnetism of the material. The value of ΔS peaks at $T_C$ of 24 K, with another shoulder near 18 K.

In order to confirm the trends in ΔS, we have also measured C in the presence of magnetic fields (see figure 1). The C(T) curves measured at different fields cross at the characteristic temperature (near 24 K). A noteworthy finding is that the peak in C(T)/T at 24 K is smeared out with the application of 10 kOe, whereas the one below 18 K undergoes a more gradual suppression with increasing magnetic fields. The ΔS(T) behavior derived from these data are also shown in figure 4a and it is to be noted that there is a good agreement between the features obtained from the two different experimental methods. Now, looking at the absolute values of ΔS, the value at $T_C$ turns out to be quite modest (about 25 mJ/ccK at $T_C$, say for a change of H from 0 to 50 kOe) when compared with those (150 – 200 mJ/ccK) for some Fe-based isostructural La compounds [10]. This could be due to the fact the saturation moment per formula unit (see figure 3) is about a factor three smaller compared to these Fe samples, apart from the possibility that the transition is not first-order in character. We also show the adiabatic temperature change (ΔT) in figure 4b, the value of which is quite significant over a wide temperature range below 50 K.

**$NdCo_9Si_4$:**

The isothermal M data, Arrott plots, ΔS and ΔT behavior are shown for this compound in figures 5 and 6 respectively. The features observed for this compound in all these data are somewhat similar to those described for the Pr compound. In particular, the strength of MCE as measured by ΔS is modest, say, for a change of H from 0 to 50 kOe, the value of which is of the order of 20 mJ/ccK at $T_C$. Correspondingly, ΔT values are also moderate.

In order to throw light on whether the magnetic transition is first order, we have carried out resistivity measurements during both warming as well as cooling cycles. We do not find any hysteresis within the detection limit in the vicinity of $T_C$ for both the samples (see Fig. 1c and 2c), which suggests that the magnetic transitions may not be first order in character. In support of this, χ(T) is also non-hysteretic at the magnetic transition. We would like to mention that the residual resistivity ratio (RRR), if normalized to the value at 40 K (to compare with the plot in Ref. 9) is close 4, which is actually larger than the specimens studied in Ref. 9. This means that the samples quality (in terms of crystallographic disorder) in our case is at least the same as in Ref. 9. [The corresponding values, if normalized to 300 K values in our case, are about 8 and 20 for Pr and Nd cases respectively].

Finally, previous literature [11, 12] indicated that there is a close relationship between MCE and magnetoresistance [MR= {ρ(H)- ρ(0)}/ ρ(0)]. In order to test this in these cases, we have obtained ρ(T) curves in the presence of different magnetic fields (see figure 1c and 2c). The values of MR thus obtained is also shown in figure 4c and 6c and it is worth noting that the magnitude of MR keeps increasing with decreasing temperature, peaking at $T_C$ like MCE. The sign of MR is essentially negative (within the limits of experimental error) in the entire temperature range of investigation expected for paramagnets and ferromagnets.

## 4.   Conclusions

The two compounds, $PrCo_9Si_4$ and $NdCo_9Si_4$, exhibit rather modest magnetocaloric effect, peaking near Curie temperature somewhat similar to magnetoresistance behavior. It appears that the magnetic transitions are not first order in these materials.

**Acknowledgement:**

The authors thank Kartik K Iyer for his help in carrying out the experiments.

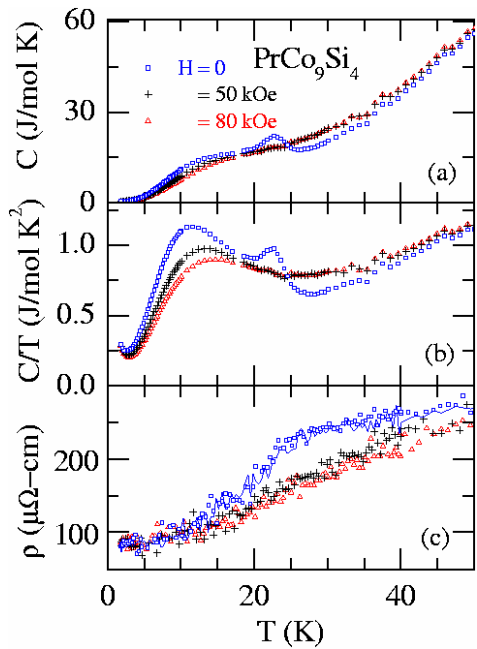

Figure 1: (color online)
(a) Heat capacity, (b) C/T and (c) electrical resistivity measured (warming cycle) in the presence of H as a function of temperature for $PrCo_9Si_4$. In (c), the data for the cooling cycle is shown as a continuous line through the data points omitting the data points for the sake of clarity.

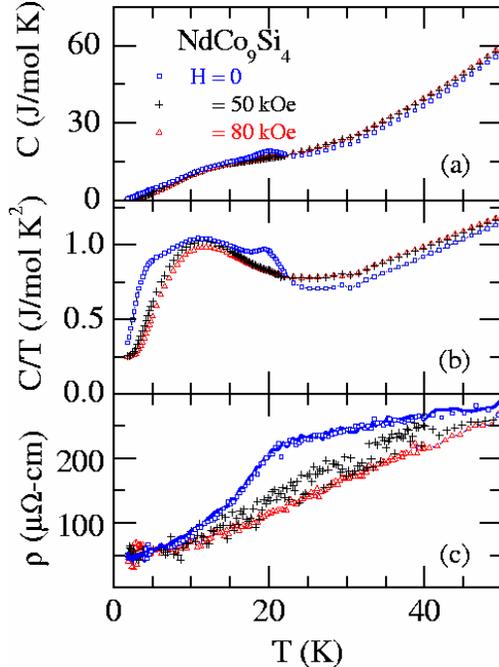

Figure 2: (color online)
(a) Heat capacity, (b) C/T and (c) electrical resistivity measured (warming cycle) in the presence of H as a function of temperature for $NdCo_9Si_4$. . In (c), the data for the cooling cycle is shown as a continuous line through the data points omitting the data points for the sake of clarity.

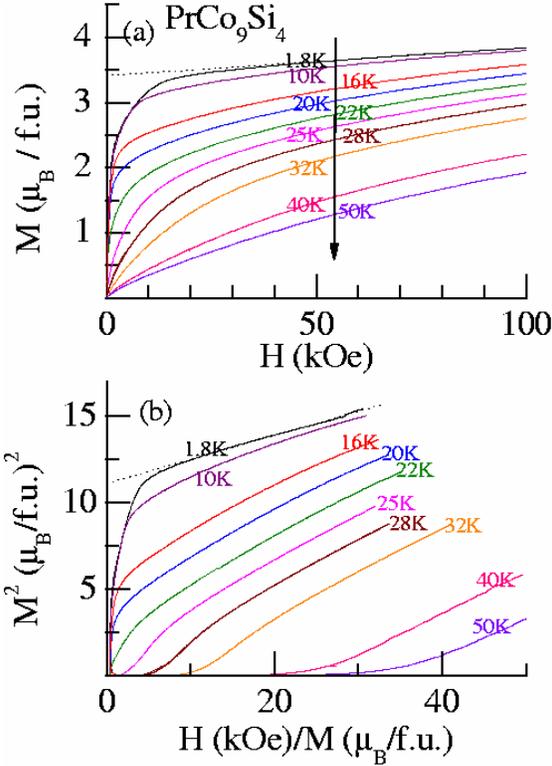

Figure 3: (color online)
(a) Isothermal remnant magnetization at various temperatures and (b) Arrott plots for $PrCo_9Si_4$. The dotted lines are drawn through the high field data at 1.8 K to highlight the extrapolated saturation moment.

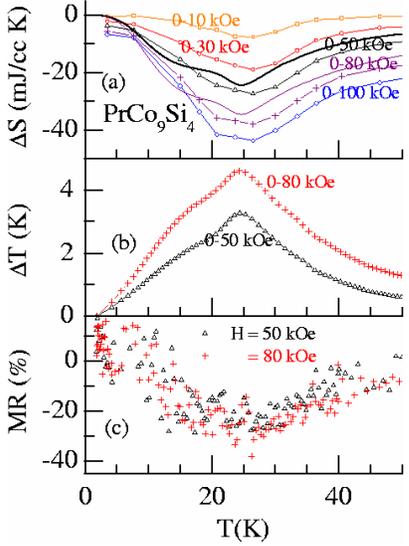

Figure 4: (color online)
(a) Isothermal entropy change, (b) adiabatic temperature change for two fields and (c) magnetoresistance, as a function of temperature for $PrCo_9Si_4$. In (a) the $\Delta S$ obtained from the heat capacity data for two magnetic fields (50 and 80 kOe) are also plotted as a continuous line (without including data points). A line is drawn through the $\Delta S$ data points obtained from M data in (a) as a guide to the eyes.

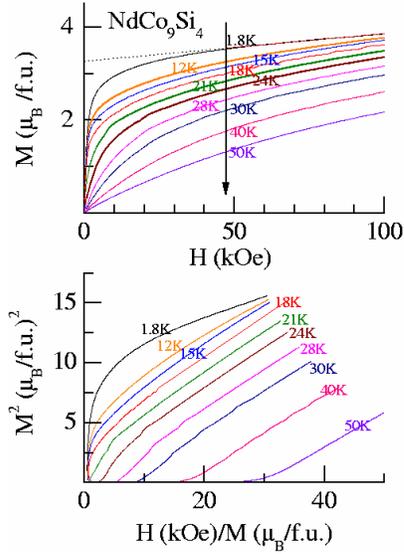

Figure 5: (color online)
(a) Isothermal remnant magnetization at various temperatures and (b) Arrott plots for $NdCo_9Si_4$. The dotted lines are drawn through the high field data at 1.8 K to highlight the extrapolated saturation moment.

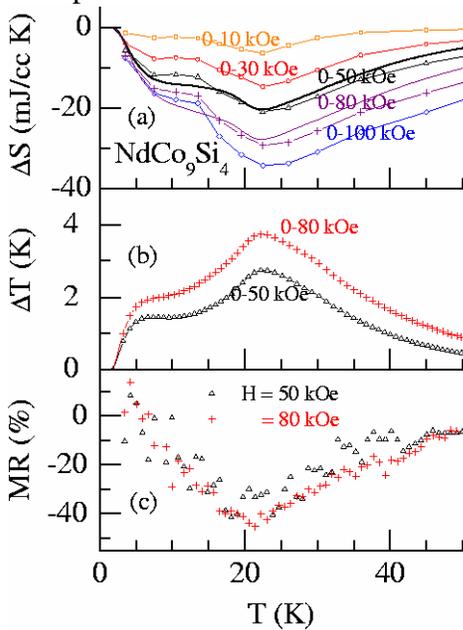

Figure 6: (color online)
(a) Isothermal entropy change, (b) adiabatic temperature change at two fields and (c) magnetoresistance, as a function of temperature for $NdCo_9Si_4$. In (a) the $\Delta S$ obtained from the heat capacity data for two magnetic fields (50 and 80 kOe) are also plotted as a continuous line without including data points. A line is drawn through the $\Delta S$ data points obtained from M data in (a) as a guide to the eyes.